\documentclass[twocolumn,prl,aps,showpacs]{revtex4}
\usepackage{graphicx}
\begin{document}
\title{Frequency down conversion through 
Bose condensation of light}
\author{Patrick Navez}
\affiliation{
Ecole Polytechnique, CP 165, Universit\'e Libre de Bruxelles, 1050 
Brussels, Belgium}
\date{\today}
\begin{abstract}
We propose an experimental set up allowing
to convert an input light
of wavelengths about $1-2\, \mu m$ into an
output light of a lower frequency.
The basic principle of operating relies on
the nonlinear optical properties exhibited
by a microcavity filled with glass. 
The light
inside this material behaves like a 2D interacting Bose
gas susceptible to thermalise and create a quasi-condensate.
Extension of this setup to a photonic
bandgap material (fiber grating) allows the light to behave 
like a 3D Bose gas leading, after thermalisation, 
to the formation of a Bose condensate.
Theoretical estimations show that a 
conversion  of $1 \,\mu m$ into
$1.5\, \mu m$ is achieved with an input pulse
of about $1\, ns$ with a peak power of $10^3 W$,
using a fiber grating containing an integrated cavity of 
size about $500\,\mu m \times 100 \mu m^2$.
\end{abstract}
\maketitle
The recent discovery of the Bose-Einstein condensate in 
atomic gas demonstrated experimentally the existence of a macroscopic 
state where a majority of particles are all in the 
same quantum state of energy \cite{Keterlee}.
This new property is of considerable interest, since it 
offers  the possibility to create an atom laser. By analogy 
with the laser, the particles move collectively to form a 
coherent wave. 
The success of this discovery is due, in part, to the ability 
of atoms to cool evaporatively in a trap. By eliminating the hotest 
atoms, some cold atoms are heated and get their energy from the 
other cold ones ensuring a thermal equilibrium. The net results are 
a cooling leading to a macroscopic population in the lowest ground state.

One question arises then: if such a thermalisation process is succesful 
for atoms, why not use it for photons? Suppose that 
a photon gas has an initial non equilibrium spectrum of frequency 
in the optical domain. If it evolves in a nonlinear medium 
to allow strong enough collision interactions between photons, then 
the thermal equilibrum spectral distribution can be reached. 

Based on simple dimensional analysis, 
we show in this paper that thermalisation of optical 
intense light is indeed possible, provided the following 
conditions are satisfied: 1) the light evolves in glass medium,
having a sufficiently low absorption, in such a way to observe 
strong enough nonlinear interactions; 2) the glass is inside 
a high finesse cavity in order to confine the photons during a time 
greater than the relaxation time towards thermal equilibrium;
3) high intensity light is required to stimulate the collision 
process by means of Bose enhancement.

Furthermore, in order to realise the (quasi-)cond\-ensation of light, 
the photon must acquire an effective mass. A two-dimensional (2D) 
massive Bose gas is created in a cavity where only 
the fundamental is considered among longitudinal
modes\cite{Chiao}.
Concerning the three-dimensional (3D) Bose gas, we propose that 
the glass is a photonic bandgap material having a band with a dispersion 
relation ressembling to that of a massive particle.  

Using the process of thermalisation or of
evaporative
cooling, two important applications using the process of thermalisation or of 
evaporative 
cooling are interesting to study: 
first, the conversion 
of a high frequency pulse of light to a pulse of lower frequencies given 
by the Bose-Einstein distribution;
second, the generation 
of a multicolor light out of a monochromatic one through an appropriate 
redistribution of the spectrum.  
We explore the possibility 
to use dielectric structures and 
multimode optical fiber gratings to realise the first 
application.


We start with the description of the 2D setup. It consists 
of a planar microcavity surrounded by two semi mirrors that 
ensure reflexion of the longitudinal component of the 
electromagnetic field. The microcavity has a transversal size 
much larger 
than the longitudinal one $L \sim \mu m$ and is filled with a 
material exhibiting significative third order optical 
non linearity together with a small absorption coefficient.
Typically, let us take the case of a glass with a linear and 
nonlinear 
refractive indexes $n_0=1.44$ and 
$n_2 \sim 3\, 10^{-20} m^2/W$ and a linear 
absorption coefficient $\alpha \sim 5\, 10^{-5} m^{-1}$. 
If the semi mirrors are perfect mirrors, then the longitudinal 
components $\vec{k}_\parallel$ of the wave vector  $\vec{k}$
take discrete values. The fundamental frequency corresponds to 
the lowest value $\omega_0= {2\pi c / \lambda_0}= {c/n_0}
|\vec{k}_{\parallel,0}|=c\pi/n_0 L$ where 
$c$ is the 
velocity of light. Selecting only this lowest value and
continuous values for the transverse components $\vec{k}_\perp$,
the energy spectrum $\epsilon(\vec{k})$ of the photon 
moving in the cavity is that of a relativistic 2D Bose gas 
with a mass $m=\hbar \pi n_0 /(Lc) \sim eV$ \cite{Chiao}:
\begin{equation}
\epsilon_{\vec{k}_\perp}=\hbar\omega_{\vec{k}}=
\frac{\hbar c}{n_0}\sqrt{\left({\pi \over L}\right)^2
+ \vec{k}_\perp^2} \simeq
\hbar\omega_0 + \frac{\hbar^2 \vec{k}_\perp^2}{2m}  
\end{equation}

The perfection of the cavity is limited by the quality of 
the semi mirror. High finesse of the order ${\cal F} \sim 10^6$ 
has been reported \cite{Kimble} with a dielectric having a thin 
layer structure made of $SiO_2$ and $Ta_2O_5$. 

When an electric field propagates inside the microcavity, a 
polarisation is created and is decomposed in three terms:
\begin{eqnarray}\label{pol}
\vec{P}(\vec{x},t)= \vec{P}_{cons}(\vec{x},t)+
\vec{P}_{dissip}(\vec{x},t)
-\vec{P}_{gain}(\vec{x},t)
\end{eqnarray}
The first term conserves the electromagnetic energy
\begin{eqnarray}
\vec{P}_{cons}(\vec{x},t)=
(\chi^{(1)} + \chi^{(3)}\vec{E}^2(\vec{x},t))
\vec{E}(\vec{x},t)
\end{eqnarray}
and describes the linear and nonlinear polarisations that contribute 
to the total effective energy hamiltonian:
\begin{eqnarray}\label{H}
H=\int_V \!\! d^3\vec{x}\,\left[ 
\frac{\epsilon}{2}\vec{E}^2(\vec{x},t)
+\frac{\mu}{2} \vec{H}^2(\vec{x},t)
+\frac{3 \chi^{(3)}}{4}\vec{E}^4(\vec{x},t)\right]
\end{eqnarray}
The susceptibility coefficients are related to those of the 
refractive index 
$n=n_0 + n_2 I$ through
$\epsilon/\epsilon_0=n_0^2$ and
$\chi^{(3)}=
\epsilon_0^2(1+ \chi^{(1)}) c n_2$.
The second term in (\ref{pol}) describes the dissipative 
losses 
$\vec{P}_{dissip}(\vec{x},t)= \alpha
\vec{E}(\vec{x},t)$.
Finally, the third 
term is present in the case photons are created by an other active 
medium inside the cavity, like a semiconductor. 
In classical physics, the presence 
of these terms allows to establish the following balance 
energy equation:
\begin{eqnarray}
{dH \over
dt}= \int_V d^3\vec{x}\,
\left(\vec{P}_{gain}(\vec{x},t)-
\vec{P}_{dissip}(\vec{x},t)\right).\vec{E}(\vec{x},t)
\end{eqnarray}
 
The quantification of the hamiltonian amounts to replacing the 
coefficients coming from each mode of the fields by the 
corresponding creation and annihilation operators 
$\hat a^\dagger_i$ and $\hat a_i$. 
Each indice $i$ labels one mode by its
wavevector components $\vec{k}$ and $\alpha$ which represents 
the polarisation state that distinguishes between TE and TM 
waves \cite{Yariv}. Developping the hamiltonian in terms of these operations, 
we make the rotating wave approximation that amounts to 
eliminating contribution which does not conserve the photon number.
One obtains \cite{Chiao}:
\begin{eqnarray}\label{Hq}
\hat H
=\sum_i
\hbar \omega_i
\hat a^\dagger_i
\hat a_i
+
\sum_{i,j,l,m}
{{\cal V}_{i,j,l,m} \over
2V}
\hat a^\dagger_i
\hat a^\dagger_j
\hat a_l
\hat a_m
\end{eqnarray}
The effective potential is assumed to be constant in 
a first approximation and is related to the scattering length 
$a$ through the relation 
${\cal V}_{i,j,l,m} = 
{4\pi a \hbar^2 \over m}$.
A comparison between the two expressions (\ref{H}) and 
(\ref{Hq}) for the hamiltonian allows to identify:
\begin{equation}
a \simeq {\hbar c^2 n_2 n_0^2\over 4\pi}\left({2\pi \over \lambda_0}\right)^3
\end{equation}
For glass and $\lambda_0=1.5 \mu m$, we obtain $a 
\sim 3\, 10^{-18}m$ much lower than the value  for atoms 
($a_{atoms}\sim 10^{-10} m$).

The frequency conversion inside the cavity proceeds in the 
following way. An initial photon pulse with a distribution 
centered in a frequency $\omega_p$ 
higher than $\omega_0$ is created inside the cavity. 
This pulse could be generated 
from an active medium or could be injected from outside. Then, 
the thermalisation process turns the initial  
distribution into  a Bose-Einstein distribution
$f(\omega_{\vec{k}_\perp})= \displaystyle
[\exp(\beta\hbar \omega_{\vec{k}_\perp} -\beta\mu) -1]^{-1}$
presenting a narrow width maximum intensity at the lowest frequency 
$\omega_0$, when the effective temperature $k_BT_{eff}=1/\beta$ 
is close to 
zero and the chemical potential $\mu$ close to $\omega_0$. 
No condensate transition is expected for a 2D  free Bose
gas but quasi-condensation is predicted for an interacting
gas.

The energy conservation imposes also that at the 
same time some higher energy photons are distributed in the tail.
If these photons escape from the cavity, then 
the process of evaporative cooling increases the 
population in the lowest levels. 
This process takes place provided that 
a lower refractive index of the coating allows the high 
transverse components of the light 
to escape from the edge of the cavity.
Assuming the minimum refractive index for the coating (air),
only modes with a wavelength greater than $\lambda=\lambda_0/n_0$ 
($1 \mu m$ for $\lambda=1.5\mu m$) 
remain confined inside the cavity.
But, as seen below, evaporative cooling is not 
necessary to convert light frequency.  

Thermalisation of the light requires 
that the photon gas remains energetically isolated. Therefore, 
the average confining time of the photon inside the cavity is much 
larger than the relaxation time towards equilibrium $\tau_{relax}$,
which is the average time between two collisions of photons 
\cite{Balescu}. The 
confining time is limited by the absorption 
inside the cavity $\tau_{abs}= 1/(\alpha c)$ and by 
the cavity finesse 
$\tau_{cav} \sim {L {\cal F} n_0 \over c}$. For the case of a 
glass, the second effect is dominant since 
$\alpha^{-1} \gg {\cal F}L$ and therefore the realisation 
condition becomes $\tau_{cav} \gg \tau_{relax}$. 

The relaxation time can be estimated 
approximatively from the kinetic theory to 
be:
\begin{equation}\label{relax}
{1 \over \tau_{relax}}=\rho \frac{c}{n_0} \sigma (1+F)
\end{equation}
and depends directly on the average photon density 
$\rho$, the velocity inside the medium $n_0 c$, the 
cross section $\sigma=4\pi a^2$ - assumed to be constant- 
and the degeneracy factor $F$ which is the average 
photon per mode. Note that if $\hbar \rightarrow 0$ then 
$a \rightarrow 0$ and no relaxation exists since in a 
classical  field system the photon does not exist. 

The factor $F$ is introduced to take into account 
the Bose enhancement which stimulates the collision process and 
diminishes the time for thermalisation.
It originates from the Uehling-Uhlenbeck quantum kinetic 
equation establishing 
the balance of modes population \cite{Balescu}.  
Third order terms in the mode 
population distribution appear in this equation and correspond
precisely to a contribution to (\ref{relax})  
quadratic in the photon density. 
Indeed, we estimate this factor by dividing
the density $\rho$ by the mode density, assuming that
the photon density is, in average, uniformously distributed
for any frequency between $\omega_0$ and $\omega_p$ and
zero otherwise. 
In this approximation, the mode
density is:
\begin{equation}
\frac{1}{L} \int_{\cal S}
\frac{d^2 \vec{k}_\perp}{(2\pi)^2} 
 = \frac{\pi}{L\lambda^2_0} \left[\left({\lambda_0 \over
\lambda_p}\right)^{2}-1\right] 
\end{equation}
where ${\cal S}$ is the area of the wavevector 
satisfying ${|\vec{k}_\perp|^2 <
\left(\frac{2\pi}{\lambda_p}\right)^{2}
-\left(\frac{2\pi}{\lambda_0}\right)^{2}}$. We
obtain
\begin{equation}
F = {L\lambda^2_0 \rho \over \pi} (\left({\lambda_0 \over
\lambda_p}\right)^{2}-1)^{-1} 
\end{equation}
Important deviation from this uniformity
occurs if the initial/final distribution is narrowly
peaked around the initial/final frequency
$\omega_p$/$\omega_0$.

A second realisation condition is that the photon number or 
power absorbed inside the medium should not exceed 
a certain amount in order not to increase considerably the 
temperature of the glass. If we authorise an increase of 
$1 K$ per pulse and since the specific heat of glass is about 
$2 J /m^3 K$, the photon density absorbed must not be greater than 
$\rho_l =2\,10^{25} photons/m^3$. If $\tau_{cav} \ll \tau_{life}$   
then the fraction absorpted is
\begin{equation}
\rho(1- \exp(-\tau_{cav}/\tau_{abs})) \simeq
\rho \tau_{cav}/\tau_{abs} \leq \rho_l 
\end{equation}
The two realisation conditions combined together impose 
the following constraints on the density:
\begin{equation}
\left[\left({\lambda_0 \over \lambda_p}\right)^2-1\right]^{1/2}
\frac{1}{2L\lambda_0 a \sqrt{\cal{F}}} \ll \rho \leq 
\frac{\rho_l}{n_0 \alpha L \cal{F}} 
\end{equation}
To satisfy these inequalities for a ratio 
${\lambda_0/ \lambda_p}=1.5$ the fidelity factor must 
obey ${\cal{F}} \leq 10^{14}$. In the realistic case of 
a dielectric with ${\cal{F}} \leq 10^{6}$ the minimum 
density 
corresponds to $\, 10^{26} photons/m^3$ or  
$2\, 10^{16} W/m^2$. 
If the input light pulse time 
is equal to $\tau_{cav}=2.5\,10^{-9}s$, we need a peak power 
of $10^6 W/cm^2$. 

The effective temperature and the chemical potential 
are estimated by 
expressing the conservation of the photon number and energy 
before and after thermalisation, neglecting the much smaller losses. 
We deduce the following balance equations:
\begin{eqnarray}
\rho =
\frac{1}{L} \int_{\cal S}
\frac{d^2 \vec{k}_\perp}{(2\pi)^2}
f(\omega_{\vec{k}_\perp})
=\frac{2}{\lambda_B^2 L}(
g_1(\kappa)+ \frac{g_2(\kappa)}{\beta \hbar \omega})
\end{eqnarray}
\begin{eqnarray}
\hbar \omega_p\rho &=&
\frac{2}{L} \int_{\cal S}
\frac{d^2 \vec{k}_\perp}{(2\pi)^2}
\hbar \omega_{\vec{k}_\perp}
f(\omega_{\vec{k}_\perp})
\nonumber \\
&=&\hbar \omega_0\rho +\frac{2}{\beta\lambda_B^2 L}(
g_2(\kappa)+ 2\frac{g_3(\kappa)}{\beta \hbar \omega})
\end{eqnarray}
where $\lambda_B=h/\sqrt{2\pi m k_B T}$ is the thermal 
wavelength and $\kappa=\beta(\hbar\omega_0 - \mu)$. 
$g_k(z)=\sum_{j=1}^\infty e^{-jz}/j^k$ are the 
Bose-Einstein functions. 
A factor $2$ takes into account the two states of 
polarisation.
The resulting thermalized pulse 
has an effective temperature of  
$T_{eff}\sim \, 10^{6} K$ and a chemical potential  
$\mu \rightarrow \hbar \omega_0$. In that situation,
most of the light is converted in the low frequency region. 
If evaporative cooling takes place, then the temperature
is lowered but some high energy photons are lost, 
limiting the conversion rate.

In the case of multimode fiber gratings, an 
all-in-one integrated structure is composed of two 
semi-mirrors and one microcavity (Fig.1).
Three typical value of 
parameters are ${\cal{F}} = 10^3$, $\alpha= 5\,10^{-2}m^{-1}$ and 
the transverse 
core section of $10^2 \mu m^2$, allowing to populate high transverse 
modes if the cladding index is close to one 
\cite{Argawal, Newport}. 
This corresponds to a minimum density of $10^{18} W/m^2$ 
and $\tau_{cav}=2.5\, 10^{-12}s$ and a peak power of $10^5 W$.
The use of such a power has
been reported in fiber grating
experiments \cite{Othonos}.

In order to increase the photon 
population in the fundamental level leading to condensation, a 3D 
cavity is needed and must allow more longitudinal 
values of the wave vector component. One possibility 
is that the cavity is itself a photonic bandgap 
material made of periodic layers, such that $\omega_0$ 
is the minimum frequency of an allowed band. In general, 
the interband interaction between particles is less 
frequent than the intraband interaction which, due to 
the difficulty to satisfy momentum-energy conservation, has 
much less probability to occur. Therefore, the thermalisation 
process is mainly achieved inside the band and generates a 
real condensate. If we assume that 
within the band the energy spectrum is of the form:
\begin{equation}
\omega(\vec{k})=
\omega_0 + \frac{\hbar (\vec{k}-\vec{k}_{\parallel,0})^2}{2 m}
\end{equation} 
then the density of modes is 
$4\pi [2 \omega_0 (\omega_p -\omega_0)]^{3/2}$.
For a longitudinal size $\tilde{L}$ of the photonic bandgap 
material, the constraints on the 
density become:
\begin{equation}
\left[\left({\lambda_0 \over \lambda_p}\right)-1\right]^{3/4}
\frac{1}{\lambda_0^{3/2} a \sqrt{{\cal{F}}\tilde{L}}} \ll \rho \leq
\frac{\rho_l}{n_0 \alpha \tilde{L} \cal{F}}
\end{equation}
For a photonic bandgap integrated inside the fiber grating 
with $\tilde{L}=10^3 L$ and $F=10^3$, 
we obtain a lower photon density of $2\,10^{16}W/m^2$, 
a much higher $\tau_{cav}=2.5\,10^{-9}s$ and a lower peak power 
of $10^3 W$
than in the case of 2D.
The effective temperature and the condensate population $\rho_0$
are estimated from the following photon number and energy 
balance equations:
\begin{eqnarray}
\rho=\rho_0 + 2\zeta(\frac{3}{2})\frac{1}{\lambda_B^3}
\\
\hbar (\omega_p-\omega_0)\rho=
2\zeta(\frac{5}{2})\frac{k_BT_{eff}}{\lambda_B^3}
\end{eqnarray}
where $\rho_0$ is the condensate density. 
We find  
$T_{eff}\sim 10^{6} K$ much lower than the critical temperature
$T_c \sim 10^{8}K$
which means that $\rho_0 \simeq \rho$. 
The dispersion frequency in  the 
resulting condensate is limited by $\Delta \omega=
1/\tau_{cav}$.

Strictly speaking, these estimations need to be validated 
by a more detailed description of the dynamical process. 
But at present, we  
prefer a model based on a simple estimation  
for two reasons. First, adequate 
kinetic equations must be formulated which take into 
account the presence of the condensate. Various models for 
atomic particle exists in the literature but no one is 
commonly accepted \cite{Cooling}. In this context, 
there is no advantage to use one more sophisticated model if 
it is not justified by experiment. Second, the particular non 
linear properties of a fiber grating at high light intensity 
are not yet well known to allow a complete characterisation 
\cite{Othonos,Kashyap}.
  
Other important non linear effects are the Raman and the Brillouin 
scatterings which can compete to convert the frequencies 
and can increase the absorption loss inside the cavity 
\cite{Argawal}. The rate at which these phenomena 
occur in an ordinary glass are estimated to be linearly proportional to  
the density with 
the constants $10^{-13} m/W$ and $6\, 10^{-11} m/W$ for 
the peak Raman-gain and peak Brillouin gain 
respectively 
\cite{Argawal}. For the light power considered, this 
corresponds to relevant rates of $10^{12}-10^{14}Hz$ which 
contribute to broaden the initial spectrum.

\begin{figure}[here]
\scalebox{0.6}{
\includegraphics{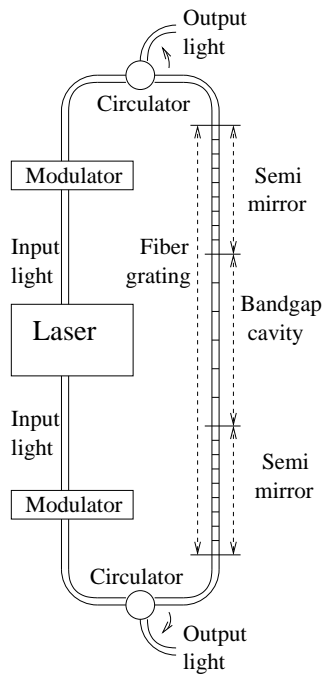}}
\caption{Setup for frequency down conversion using fiber grating}
\end{figure}

Another important question is how to inject the light inside 
the cavity. If we exclude the use of an active medium, the light 
can be introduced transversaly on the edge, by coupling the cavity 
medium with optical fiber or with a prism \cite{Stegeman}. 
In the case of fiber gratings, the spatial profile of the 
pulse should be designed in order to inject a maximum 
percentage of the initial intensity. This profile is 
determined by noticing the time reversal process describes 
an initial density distribution of light 
inside the cavity, evolving towards a final density distribution
outside the cavity. This final distribution
is simulated according to the dynamical model used which, 
in the case of low intensity, is governed by the linear 
classical Maxwell 
equations.  Since  the dynamic is time 
reversal invariant, this final density distribution 
gives precisely the initial profile to be 
used in order to confine all the light inside the cavity and has 
a size of the order of $\tau_{cav}c/n_0$. Fig. 1 depicts a possible 
setup. Two initial pulses of light generated by a laser  are modulated 
to get the initial profile and enter coherently in the fiber grating. 
After the frequency conversion takes place, 
the resulting light distribution 
leaves the cavity and gets out by means of the circulators.     

In conclusion, we show the feasibility to convert the 
frequency of a light inside an optical nonlinear cavity. 
The analysis is based on an estimation of the collision 
rate between photons in a glass, leading to a thermalisation 
and a high condensed population in the fundamental mode 
of the cavity.

{\bf ACKNOWLEDGEMENTS}

P.N. thanks Prof.
L.A. Lugiato, 
N.J. Cerf, S. Massar and the members of the Service d'Optique 
et Accoustique of U.L.B.  
for helpful discussions. 
P.N. acknowledges funding by the European Union under the 
project EQUIP (IST-FET programme) and by the  
`` Action de Recherche Concert\'ee '' of the Communaut\'e 
Fran\c caise de Belgique.

\end{document}